\documentclass[aps,pra,twocolumn, groupedaddress]{revtex4-2}

\usepackage{graphicx}
\usepackage[T1]{fontenc}
\usepackage[utf8]{inputenc}
\usepackage{xcolor}

\newcommand{\beq}{\begin{eqnarray}}
\newcommand{\eeq}{\end{eqnarray}}

\newcommand{\Dp}[2]{\frac{\partial #1}{\partial #2}}
\newcommand{\DDp}[2]{\frac{\partial^2 #1}{{\partial #2}^2}}
\newcommand{\dd}{\ensuremath\mathrm{d}}
\newcommand{\half}{\frac{1}{2}}
\begin{document}


\title{Quantum and classical branching flow in space and time}


\author{Jakub Šťavina}
\email[]{stavina.jakub@gmail.com}
\affiliation{Department of Physics and Astronomy, The University of Manchester, Manchester M13 9PL, England, United Kingdom}


\author{Peter Bokes}
\email[]{peter.bokes@stuba.sk}
\affiliation{Faculty of Electrical Engineering and Information Technology, Slovak University of Technology, 81219 Bratislava, Slovak Republic}
\date{\today}

\begin{abstract}
Branching flow -- a phenomenon known for steady wave propagation in two-dimensional weak correlated random potential is also present in the time-dependent Schr\"{o}dinger equation for a single particle in one dimension, moving in a fluctuating random potential. 
We explore the two-dimensional parameter space of this model using numerical simulations and identify its classical regions, 
where just one classical parameter is sufficient for its specification, and its quantum region, where such a simplification is not possible. 
We also identify region of the parameter space where known analytical results of a classical white-noise model are relevant. Qualitative behavior 
of quantum and classical particle dynamics is discussed in terms of branching time scale and a new time scale related to particle's kinetic energy.

\end{abstract}


\maketitle

\section{Introduction\label{sec-intro}}



In 2001 Topinka {\it et al.}\cite{Topinka2001} discovered an interesting behavior of electrons entering a weakly perturbed
2D degenerate electron gas through a narrow constriction. The electronic flow exhibited
branching strands on a spatial scale larger than the correlation length of the perturbing potential.
The phenomenon has been explained as a classical flow of particles in 2D weak short-ranged correlated random potential 
whose trajectories are repeatedly focused into caustics~\cite{Topinka2001,Kaplan2002}. 
The flow has several universal statistical features~\cite{Kaplan2002,Metzger2010} such as a simple scaling for the branching time $t_\mathrm b$ 
with its universal dependence on weak random potential amplitude strength $t_\mathrm{b} \sim v_0^{-2/3}$.
Behavior similar to branching flow has been known earlier as an explanation for the appearance of large surface waves in oceans 
due to random topography of the ocean floor playing the role of weak random potential~\cite{Berry2007, Degueldre2015}. 
Motivated by the strongly anisotropic
structure of the oceans' floor caused by its geological history, the classical model describing the
branching flow has been generalized for 2D anisotropic random potential with two different correlation lengths~\cite{Degueldre2017}.
Importance of wave decoherence and smoothing of ray dynamics has been stressed for the appearance of the branches with high intensity~\cite{Berry2020}.
Recently, Patsyk {\it et al.}~\cite{Patsyk2020} observed branching flow of laser light in thin soap bubbles which once again
demonstrated the statistical characteristics of branching flow. Their experiment was so simple to understand and its
results so beautiful that the effect found its way as one of the problems for the 34th International
Young Physicists' Tournament in 2021~\cite{IYPT}. An overview and explanation of the branching flow phenomenon for
a broader audience has been recently published in~\cite{Heller2021}.

In our work we explore the signatures of branching flow phenomenon in the time-dependent Schr\"{o}dinger
equation for a single particle in one dimension, moving in a fluctuating random potential. This is motivated
by the similarity of the paraxial approximation for static 2D wave equation used for
the description of branching flow~\cite{Patsyk2020} to the time-dependent Schr\"{o}dinger equation.
In our models the role of the static 2D correlated disorder is taken by a
correlated fluctuating disordered potential $V(x,t)$. In contrast to the typically considered static disordered potential
in two dimensions, the two variables $x$ and $t$ are quite distinct and the correlation of the potential energy
is characterized by two independet parameters -- the correlation length $\lambda$ and the correlation time $\mathcal T$.
This makes our problem mathematically similar to the anisotropic random potential considered
by Degueldre {\it et al.}\cite{Degueldre2017}.

We study three levels of theory for the model: (1) a quantum particle in a fluctuating disordered potential, (2) a classical
particle in a fluctuating disordered potential and (3) a classical particle affected by a random force with white noise
correlation function. The first two are studied numerically and their mutual comparison leads to identification of a transition
between quantum-mechanical and classical dynamics in an identically defined stochastic potential. This transition is also manifested
through a reduction of the two-parametric system to a system with a single classical parameter. For the third case,
which is a physically motivated approximation of the second one, we discuss available analytical results,
and compare them to numerical results for the first two cases. Based on this comparison we explore the limitations of
the analytical formulae that can be used also for discussion of branching flow in wider range of physical systems.

\section{Quantum particle in fluctuating disordered potential \label{sec-quantum-model}}

We want to describe the dynamics of a quantum particle with mass $m$ moving in one spatial dimension within 
a fluctuating disordered potential $V(x,t)$.
The potential is a result of certain stationary stochastic process with zero mean and prescribed correlation function
\beq
        \langle V(x'+x,t'+t) V(x',t') \rangle = V_0^2 s_x(x/\lambda) s_t(t/\mathcal T),  \label{eq-VV}
\eeq
where $V_0^2$ gives the variance of the potential energy fluctuations,
$\lambda$ its correlation length in spatial dimension and $\mathcal T$ its correlation time.
The functions $s_x(z)$ and $s_t(z)$ are even, equal to 1 at $z=0$ and decay quickly to zero.
In our work we use Gaussians for both $s_x(z) = s_t(z) = \exp(-z^2/2)$. It has been claimed
in the past that the typical behavior of the observed branching flow patterns is rather insensitive to the
precise form of these functions~\cite{Topinka2001,Patsyk2020}.

In general, the averaging indicated by the brackets $\langle ... \rangle$
corresponds to ensemble average over many realizations of the potential.
In our numerical simulations we consider more restrictive ensemble
of potentials, where 
Eq.~\ref{eq-VV} is satisfied by requiring each potential in the ensemble to fulfill 
$
	( L T )^{-1} \int_0^L \dd x' \int_0^T \dd t' V(x'+x,t'+t) V(x',t') 
	= V_0^2 s_x(x/\lambda) s_t(t/\mathcal T)
$
for finite domains of integration $L$ and $T$. This can be understood 
as averaging over potentials generated by a shift in space and time from a single realisation only.

The correlation length implies a relevant spatial scale for the problem
which we use as a unit of length, $x/\lambda \rightarrow  x$.
On the other hand, as a unit of time we choose a quantum time scale associated
with a distance $\lambda$, $t / [(m \lambda^2)/\hbar] \rightarrow t$.
Using these \emph{quantum units} (q.u.) the Schr\"{o}dinger equation for the particle is
\beq
    i \Dp{\psi}{t} =
   - \frac{1}{2} \DDp{\psi}{x}  + v_0 \xi(x,t/\tau) \psi  .  \label{eq-SE}
\eeq
where $v_0 = (m\lambda^2/\hbar^2) V_0$, $\tau = \mathcal T / [(m \lambda^2)/\hbar]$ and
$\langle \xi(x'+x,t'+t) \xi(x',t') \rangle  = s(x) s(t)$.
The use of quantum units results in two parameters $v_0$ and $\tau$ that
define the fluctuating disordered potential. The meaning of $v_0$ is the ratio of the mean square fluctuation
of the potential to the estimate of the ground state energy of a particle in a well of length $\lambda$, the meaning
of $\tau$ is the ratio of the fluctuating potential's correlation time to the estimate of the period of oscillation in time
of the phase of a particle in the ground state of the well of length $\lambda$.

The expression $\tilde v_0 = v_0 \tau^2 $, when cast into S.I. units, gives $V_0 {\mathcal T}^2 / (m \lambda^2)$, i.e.
an expression which is independent of Planck's constant. Hence, this combination remains finite in the classical
limit and represents a single parameter of the system in the classical regime, which we confirm in the next section.

On the other hand, both $\tau$ as well as $v_0$ in S.I. units contain the Planck's constant and in the classical limit
$\hbar \rightarrow 0$ we find $\tau \rightarrow 0$ and $v_0 \rightarrow \infty$. In view of their above interpretation
using a quantum particle in a potential well, $v_0 \leq 1$ and at the same time $\tau \ge 1$ should imply non-classical
behavior.
We note that $\tau \rightarrow \infty$ results in a static disorder, a situation studied in the theory
of strong localization in 1D~\cite{Markos2006}. On the other hand, the limit $v_0 \rightarrow 0$ corresponds
to a free quantum particle.

Formally, Eq.~\ref{eq-SE} is equivalent to the paraxial approximation of a stationary wave equation in 2D~\cite{Patsyk2020}.
In the case of light propagation in a thin soap film studied by Patsyk et al.~\cite{Patsyk2020},
the correspondence is obtained using the following identifications:
$v_0 = l_c^2 k_0^2 \bar{n}^2 u_0$ and $\tau = (k_0 \bar{n} l_c)^{-1}$,
where we use the notation from the original paper~\cite{Patsyk2020}:
$l_c$ is the correlation length of the 2D static disordered potential,
$k_0$ wave number of the light in the vacuum and
$u_0 = 0.5\sqrt{\langle n^4_\mathrm{eff}\rangle/\bar{n}^4-1}$ with $\bar{n}$ and
$n_\mathrm{eff}$ being the average and effective refractive index. Through this
expressions we find that, for example, for Fig.3a in \cite{Patsyk2020} the corresponding values of our parameters are
$v_0  = 4.42 \times 10^5 \gg 1$ and $\tau = 3.36 \times 10^{-4} \ll 1$ so that we are safely in a classical regime.
The classical parameter $\tilde v_0$ attains for this particular example value $\tilde v_0 = v_0 \tau^2 \approx 0.05$.

Similarly, we can relate the model defined by Eq.~\ref{eq-SE} to the branching flow observed in electronic flow through
quantum point contact in 2D electron gas~\cite{Topinka2001}. The paraxial approximation was not used in this work.
Instead, the authors used a full 2D propagation for their simulations~\cite{Shaw2002}.
Nonetheless, to place their physical realization of branching flow into parameter space of our model we use
the correspondence between the paraxial approximation to the time-independent Schr\"{o}dinger equation in 2D
and Eq.~\ref{eq-SE} that results in simple mapping between the parameters of both models:
$v_0 = (m_e l_c^2/\hbar^2) V_0 \approx 0.83 \sim 1$ and $\tau = (k_F l_c)^{-1} \approx 0.22 \sim 1$,
where $k_F$ is the electron's Fermi wave vector, correlation length of the 2D static disordered potential is $l_c$
and the square root fluctuation of the potential $V_0$.  These values of $\tau$ and $v_0$ indicate quantum regime,
while the classical parameter $\tilde v_0 \approx 0.04$ is similar to the case of light propagation
in soap films~\cite{Patsyk2020}.

\section{Classical and white noise models \label{sec-classical-model}}

The classical equation of motion corresponding to the dynamics of quantum particle from the previous section is
\beq
        \ddot{x} &=& - v_0 \Dp{~}{x} \xi(x,t/\tau)  \label{eq-newton}
\eeq
Here, however, one may reduce the two parameters $v_0$ and $\tau$ to one by rescaling the units
of time once again according to $t/\tau \rightarrow t$, with the result of having a single classical
parameter $\tilde{v}_0 = v_0 \tau^2$ identified in Sec.~\ref{sec-quantum-model} based on dimensional analysis.
This parameter sets the strength of the fluctuating force
$f(x,t) = - \tilde v_0 \Dp{~}{x} \xi(x,t)$  on the right hand side of a rescaled version of Eq.~\ref{eq-newton}.

To obtain a couple of analytical results we use an ad-hoc approximation to the fluctuating force
$f(x,t)$ in the form of a spatially uncorrelated random force $f(t)$ with $\langle f(t) \rangle = 0$
and $\langle f(t) f(t') \rangle =  \tilde{v}_0^2 s(t - t')$.
In other words, we assume that this force's time correlation function inherits the correlation time $\tau$
($\tau = 1$ in the rescaled units) of the original potential $V(x,t)$ from Eq.~\ref{eq-VV}.
Within this approximation it is possible to show that the average kinetic energy of a particle is given by the expression
\beq
        \langle e_k(t) \rangle &=& e_k(0) +  {\tilde v_0}^2  \sqrt{\pi} \int_0^{t/\sqrt{2}} \label{eq-ek-exp}
        \mathrm{erf}(z) \dd z .
\eeq

Even simpler model for the fluctuating force, used by several studies in the past~\cite{Kaplan2002,Degueldre2017},
is the white noise model of the fluctuating force
\beq
f_\mathrm{w.n.}(t) = \gamma \Gamma(t), \quad
\langle \Gamma(t) \Gamma(t') \rangle =  \delta( t - t')   \quad
\label{eq-white-noise-1}
\eeq
for which the average kinetic energy of a particle is given by a simple expression
\beq
     \langle e_k(t) \rangle  = e_k(0) + \gamma^2 t  . \label{eq-ek-wn}
\eeq
Identical linear growth in time is obtained also from the long-time behavior of the average kinetic energy 
from Eq.~\ref{eq-ek-exp} for $\gamma^2={\tilde v_0 }^2 \sqrt{\pi/2}$. In other words, the white noise model 
is an approximation to the random force model valid for times $t \gg \tau = 1$. 
Hence, its predictions need to be limited to the time scales large compared to the fluctuating potential's correlation time at best.

The second quantity of interest for which an analytical result for the white noise model is available~\cite{Degueldre2017}
is the average square of the particle's displacement $\sigma_x^2(t) = \langle [x(t) - x(0)] ^2 \rangle$,
\beq
   \sigma_x^2(t) = \dot x(0)^2 t^2 + \frac{2}{3} \gamma^2  t^3 .  \label{eq-x2-wn}
\eeq
From the two analytical results (\ref{eq-ek-wn}) and (\ref{eq-x2-wn}) we can obtain estimates
of two different time scales encountered in the dynamics of particle in fluctuating disordered potential.
In their calculation we will assume that the initial average
kinetic energy is negligible compared to the kinetic energy attained by the particle in the course of its motion.

The branching time $t_\mathrm{b}$ is defined as the time for which the average displacement Eq.~\ref{eq-x2-wn}
is equal to the correlation length $\lambda$ ($\lambda=1$ in our present units) of the disordered potential, 
\beq
        t_\mathrm{b} = \left( \frac{3}{2 \gamma^2} \right)^{1/3} = \left( \frac{9}{2\pi} \right)^{1/6} \tilde{v}_0^{-2/3}
\eeq
As noted earlier, the validity of the white noise model is limited at best to time scales $t_\mathrm{b}  \gg \tau =1$ which
results in the requirement $\tilde v_0 \ll \left( 9 / (2 \pi) \right)^{1/4} \approx 1.09$.

Since the kinetic energy in the random force model increases with time, we also define time $t_{e}$
being the instance when the average kinetic energy is equal to the square root dispersion of the potential energy fluctuations,
\beq
        t_{e} = \frac{\tilde v_0}{\gamma^2} = \frac{1}{\sqrt{\pi/2} \tilde v_0} .
\eeq
If this estimate could be used
also within the fluctuating potential model then it would indicate that beyond this time scale the particle
will propagate over long distances and  it would not be bound within a local minimum of the potential.

The usefulness of these formulae rests on the validity of the white noise model for the fluctuating random potential.
To explore this we have performed numerical simulations for both the classical and the quantum system
with fluctuating disordered potential that are discussed in the next section.

\section{Numerical simulations}

In our numerical simulations we used quantum units introduced in section~\ref{sec-quantum-model} for quantum as well
as the classical dynamics. The analytical results from the previous section using the quantum units are
\beq
     \langle e_k(t) \rangle  &=& e_k(0) + \sqrt{\frac{\pi}{2}} v_0^2 \tau t  \label{eq-ek-wn-qu} \\
     \sigma_x^2(t) &=& \dot x(0)^2 t^2 + \frac{\sqrt{2\pi}}{3} v_0^2 \tau  t^3 .  \label{eq-x2-wn-qu} \\
     t_\mathrm{b} &=& \left( \frac{9}{2\pi} \right)^{1/6} v_0^{-2/3} \tau^{-1/3} \label{eq-tb-qu} \\
     t_{e} &=& \frac{1}{\sqrt{\pi/2} v_0 \tau}  . \label{eq-tv0-qu}
\eeq

 Realizations of the functional form of the fluctuating disordered potential $\xi(x, t)$ in Eq.~\ref{eq-SE} are generated
 using uniform random distribution for the phases of its Fourier series  $\mathcal{F}[\xi](k_n,\omega_m)$ in the domain $(x,t) \in (0,L) \times (0,T)$.
 The magnitude of the Fourier coefficient is determined by the Fourier coefficient $\mathcal{F}[s](k_n)$ 
 of the correlation functional form $s(z)$ (Eq.~(\ref{eq-VV})) 
 by the expression $|\mathcal{F}[\xi](k_n,\omega_m)| = \sqrt{|\mathcal{F}[s](k_n)| |\mathcal{F}[s](\omega_m)|}$.
 In this way the ensemble of generated potentials fulfills correlation function Eq.~\ref{eq-VV} by construction.
(See supplementary information ~\cite{supplement} for more details).

\begin{figure}[t]
	\includegraphics[width=0.5\textwidth]{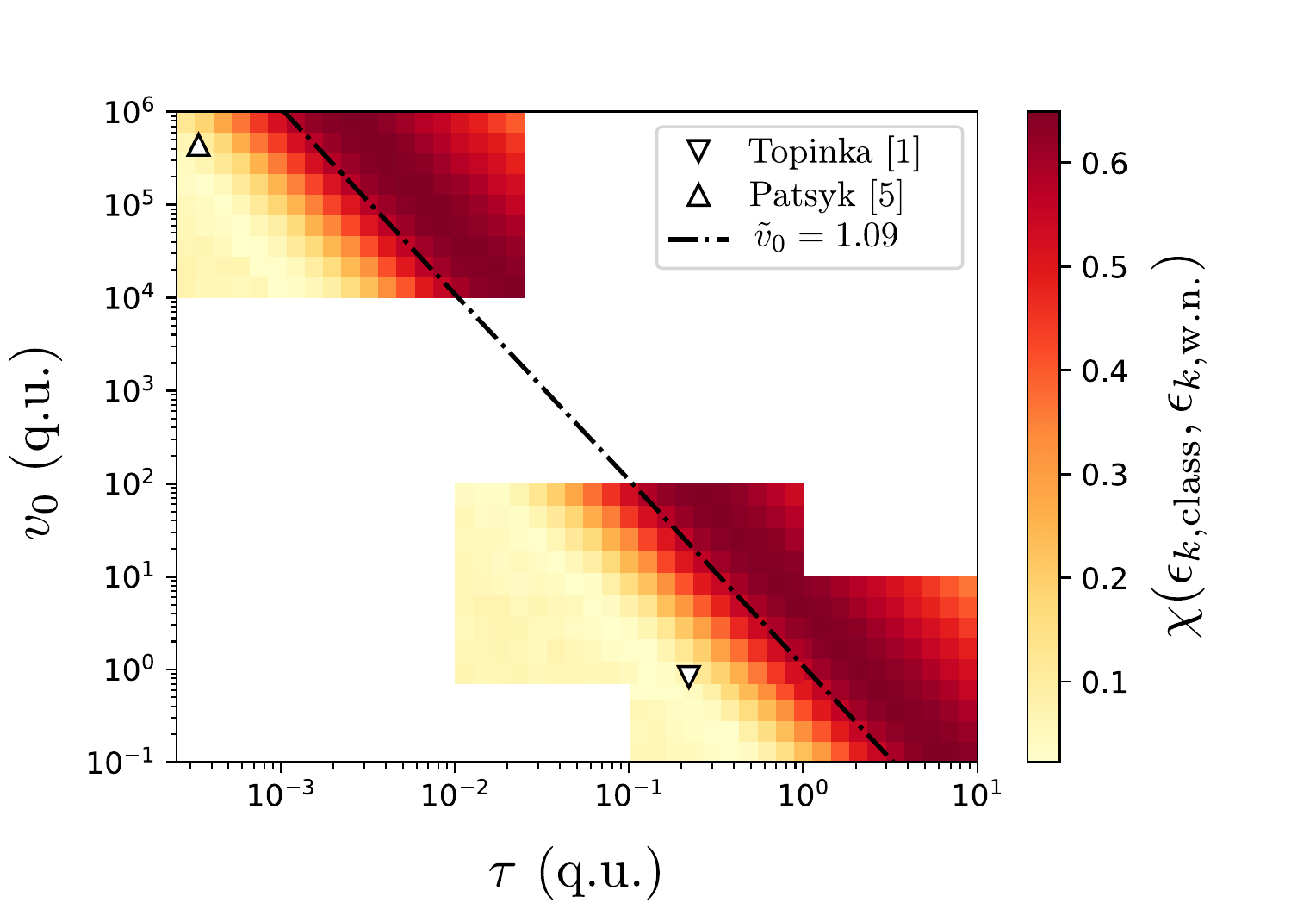}
        \caption{
        The kinetic energy difference indicator for numerical solution of the classical model 
	(Eqs.~\ref{eq-newton}) and analytical results for the white noise models shown in color 
	within the considered parameter space $\tau \times v_0$.
	Both models for different values of parameters are equivalent along the lines 
	$\tilde v_0 = v_0 \tau^2 = \mathrm{const}$. In the left bottom part of the parameter space, 
	left of the dashed line $\tilde v_0 = 1.09$ the results for numerical solution and the white noise model agree. 
	Triangular symbols show the positions of parameters for systems experimentally studied
        by Topinka et al.~\cite{Topinka2001} and Patsyk et al.~\cite{Patsyk2020}.
	\label{fig-parameter-space-cl}}
\end{figure}

\begin{figure}[t]
\includegraphics[width=0.5\textwidth]{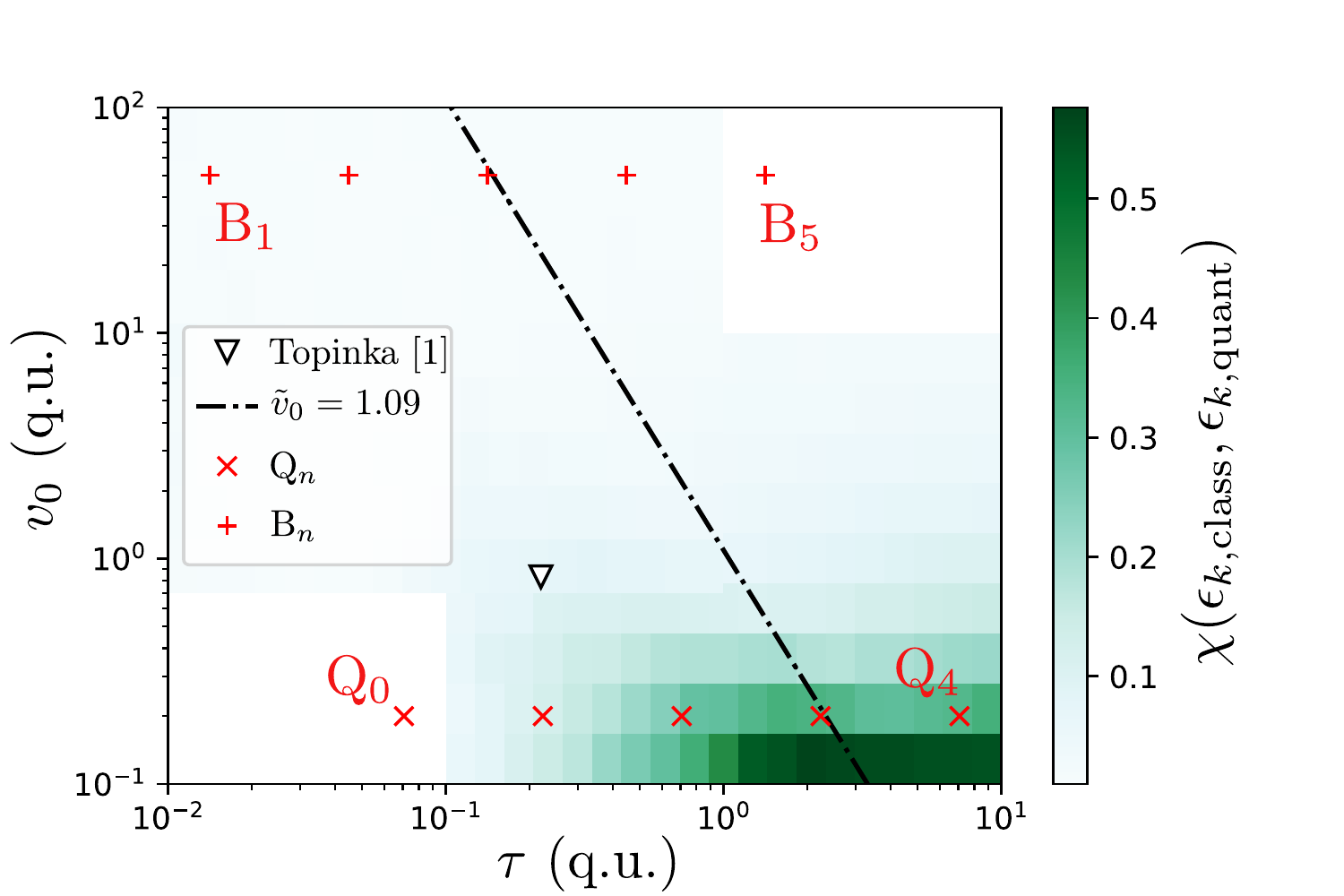}
        \caption{The kinetic energy difference indicator for the classical and quantum behavior.
        Points B$_n, n = 1,...,5$ and Q$_n, n = 0,...,4$ show the position of classical and quantum regimes
        which we analyze as the fluctuating potential becomes slower (increasing $\tau$).
        The triangular symbol gives the parameters of the system corresponding to that studied by Topinka et al.~\cite{Topinka2001}.
	\label{fig-parameter-space} }
\end{figure}

Samples of potentials have been generated on a spatial domain $x\in (0,100)$  with typically $N=8000$ samples and on a time
domain $t \in (0,5t_\mathrm{b})$ with typically $M=30000$ samples for the purpose of parameter space scanning presented in 
Figs.~\ref{fig-parameter-space-cl} and \ref{fig-parameter-space}. For longer runs (larger value of $T$) the time step
$\Delta t$ was chosen such as to keep the ratio $\Delta t / (\Delta x^2) = 1$, where $\Delta x = L/N$, 
which resulted in satisfactory stability of numerical integration of both classical as well as quantum equations of motion. 

The classical simulations were performed using second order St\"{o}rmer -- Verlet integrator for
a set of $4000$ initial positions uniformly distributed over the whole spatial domain.
The initial velocity was set to zero. From a sample of 104 
random fluctuating potentials, each with all the 4000 initial positions
we calculated the time evolution of the average kinetic energy
$\epsilon_{k,\mathrm{class}}(t) = \langle \frac{1}{2} \dot x(t)^2 \rangle$,
and the time evolution of the root mean square displacement of the particle's coordinate
$\sigma_{x,\mathrm{class}}(t) = \sqrt{ \langle (x(t) - x(0))^2  \rangle }$.

The quantum simulations were performed using the split-step Fourier method which utilizes Suzuki–Trotter formula for
the time evolution operator. The potential energy term of the Hamiltonian was propagated in the real space
and the kinetic energy term was propagated in the Fourier space.
We have used two kinds of initial conditions: a single Gaussian wave packet with unit width localized in the center
of the simulation domain ($x=L/2$) for the calculation of the mean square displacement of particle's coordinate
and a constant amplitude (``plane wave'') over the whole simulation domain for the calculation of
the average kinetic energy and the scintillation index (see below).
The initial kinetic energy of the plane wave state is negligible,
whereas the localized initial condition in the form of Gaussian of unit width has
the initial kinetic energy $\epsilon_{k,\mathrm{quant}}(0) = 1/4$. 
The root mean square displacement for the Gaussian initial condition is 
$\sigma_{x,\mathrm{quant}}(0) = 1/\sqrt{2}$. Similarly to the classical simulations, for 104 samples of random potentials,
we calculated the average kinetic energy 
$\epsilon_{k,\mathrm{quant}}(t) = - \langle \int \psi^*(x,t) \half \DDp{~}{x} \psi(x,t) \dd x \rangle$
for the plane wave initial conditions and the root mean square displacement of particle's coordinate
$\sigma^2_{x,\mathrm{quant}}(t) = \langle \int |\psi(x,t)|^2 (x - L/2)^2 \dd x \rangle$ for the initial
condition localized at $x=L/2$, in the center of the simulation box. 
Due to the finite simulation domain the displacement calculation does not lead to useful results 
when the amplitude of the wave function is noticeable at the box boundaries $x=0$ or $x=L$.
Hence, this quantity will be taken into account only for a limited time of simulation.
In contrast, the average kinetic energy is insensitive to the boundary condition due to the fluctuating nature of the potential. 
This is also the reason why plane wave initial condition is suitable for its calculation.

We have performed the above described calculations for a wider region of models' parameter space
which is shown in Fig.~\ref{fig-parameter-space-cl}.
To quantify differences between the time-dependence of the average kinetic energy calculated for two different models
we use an indicator $\chi$ attributed to two functions $a(t)$, $b(t)$ 
by the formula $\chi(a,b) = \sqrt{\sum_{i=1}^M \left( 1-a(t_i)/b(t_i) \right)^2 / M }$ where $\left\{ t_i \right\}_{i=1}^M$ is 
the used time discretization.
From  Fig.~\ref{fig-parameter-space-cl} we can clearly see that approximating the fluctuating random potential
by the white noise model does indeed describe the behavior well if the condition $\tilde v_0 \ll 1$ is fulfilled.
The figure also demonstrates that the classical models depend only on the classical parameter $\tilde v_0 = v_0 \tau^2$ 
and not on the two parameters $v_0$ and $\tau$ separately. In contrast, comparison of the numerical simulations of the classical
and the quantum particle in Fig.~\ref{fig-parameter-space} shows that quantum
model depends on two separate parameters in the non-classical region of the parameter space, $\tau > 1$ and $v_0 < 1$.
This also confirms the criteria for quantum behavior stated in Sec.~\ref{sec-quantum-model}.

\begin{figure}[t]
\includegraphics[width=.5\textwidth]{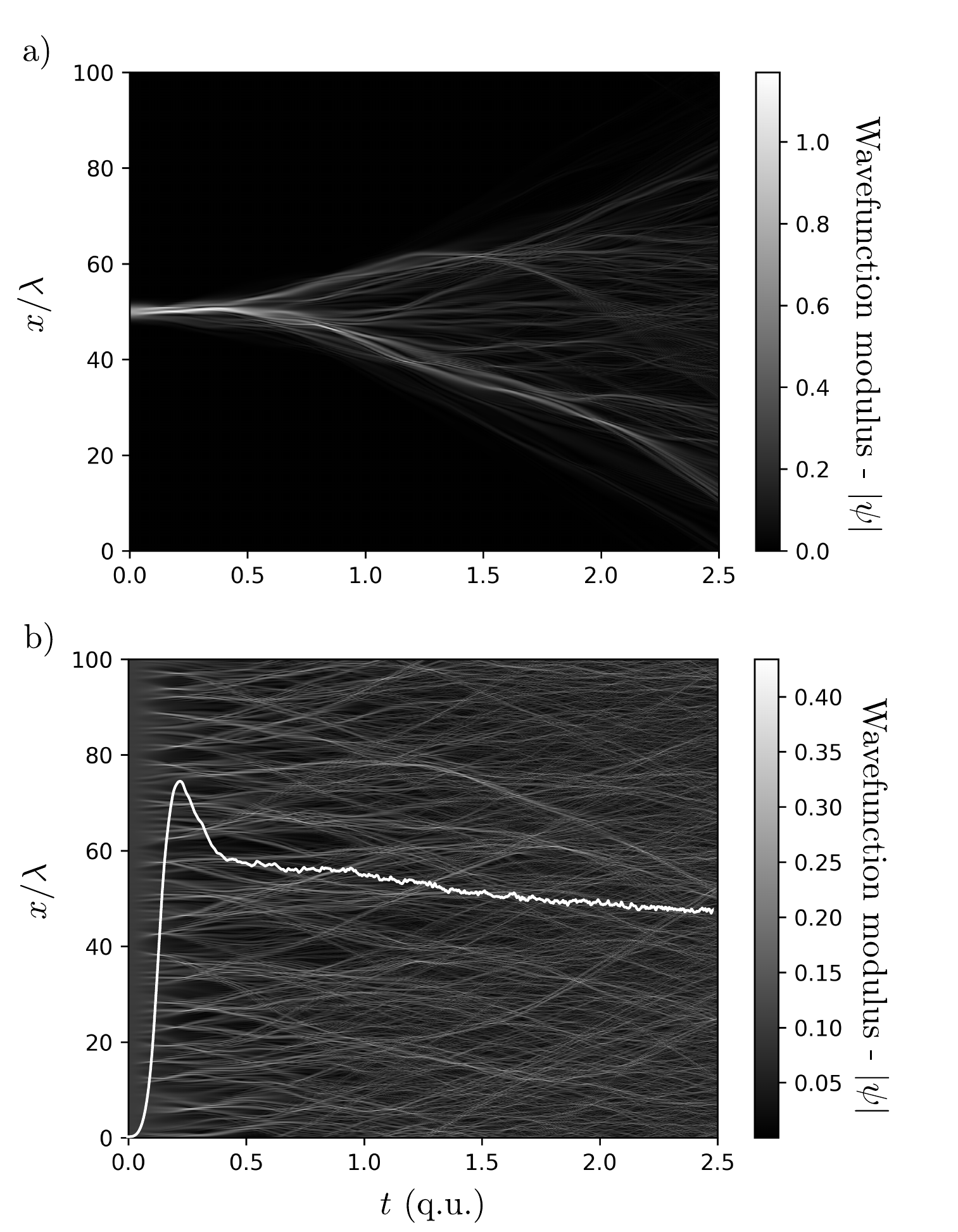}
        \caption{Branching flow pattern of the propagated wave function amplitude for
        $v_0 = 50$ and $\tau \approx 0.05$ (B$_2$) when the initial condition is
        a) a localized Gaussian wave packet and b) a constant. Over the wave function amplitude
        in b) we show the scintillation index having a pronounced maximum close to the branching time
        $t_\mathrm b = 0.22$. 
	Figures corresponding to other B$_n$/Q$_n$ points can be found in supplementary information ~\cite{supplement}. 
	\label{fig-amplitudes} }
\end{figure}

The wave function modulus $|\psi(x,t)|$ for the specific quantum simulation
with $\tau \approx 0.05$ and $v_0 = 50$ ($\tilde v_0 = 0.125$) is
shown in Fig.~\ref{fig-amplitudes}. Both localized and plane wave initial conditions result in a typical
branching flow patterns for this value of parameters. The scintillation index,
$S(t) = {\langle(\int |\psi(x,t)|^4 \dd x \rangle}/{\langle \int |\psi(x,t)|^2 \dd x \rangle^2} - 1$~\cite{Patsyk2020}
is shown by the white curve on top of the amplitude for plane wave initial condition and it exhibits a maximum close
to the branching time $t_{\mathrm b}$. 
We note that in quantum regime this feature of the scintillation index disappears, as it is demonstrated in 
Figs. 7-12 in the supplementary information~\cite{supplement}.

\begin{figure}[t]
        \includegraphics[width=.5\textwidth]{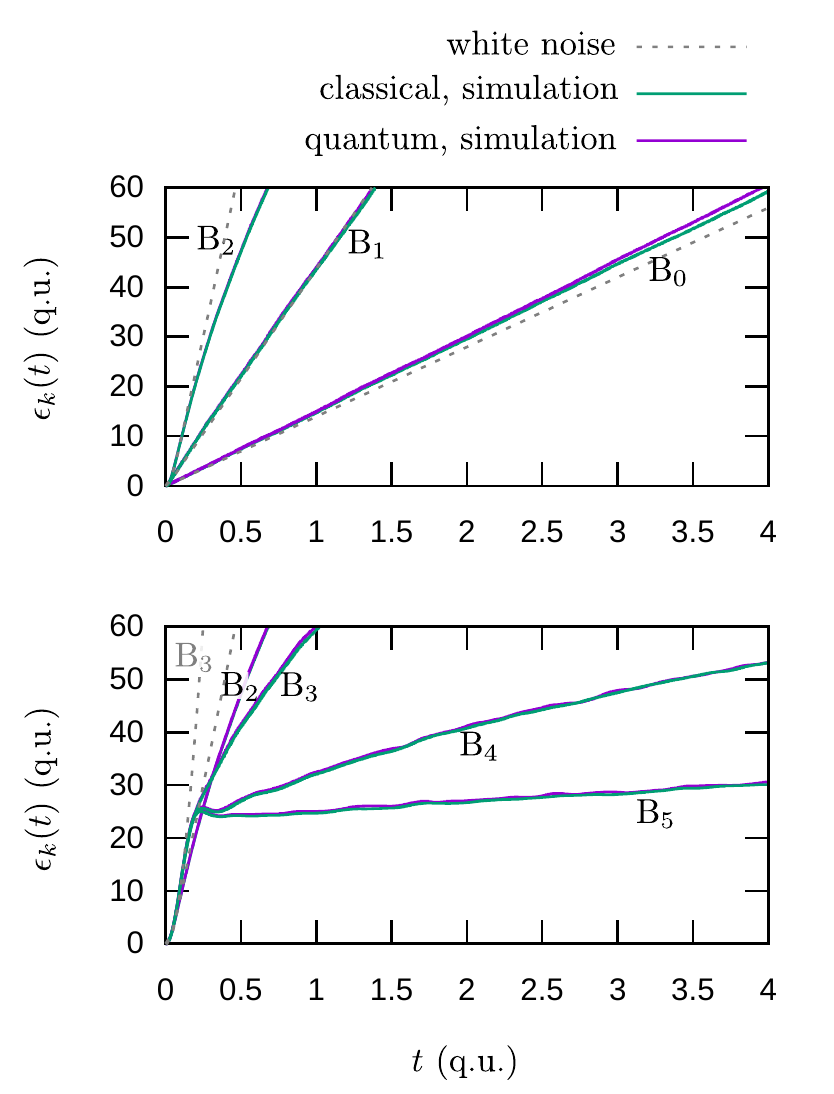}
        \caption{Dependence of average kinetic energy on time for the
        series of ``classical'' points B$_n$ in the parameter space.
        Up to $n=2$ the white noise model is in agreement with both simulations,
	beyond $n=2$ the behavior changes qualitatively. \label{fig-ek-B} }
\end{figure}

\begin{figure}[t]
        \includegraphics[width=.5\textwidth]{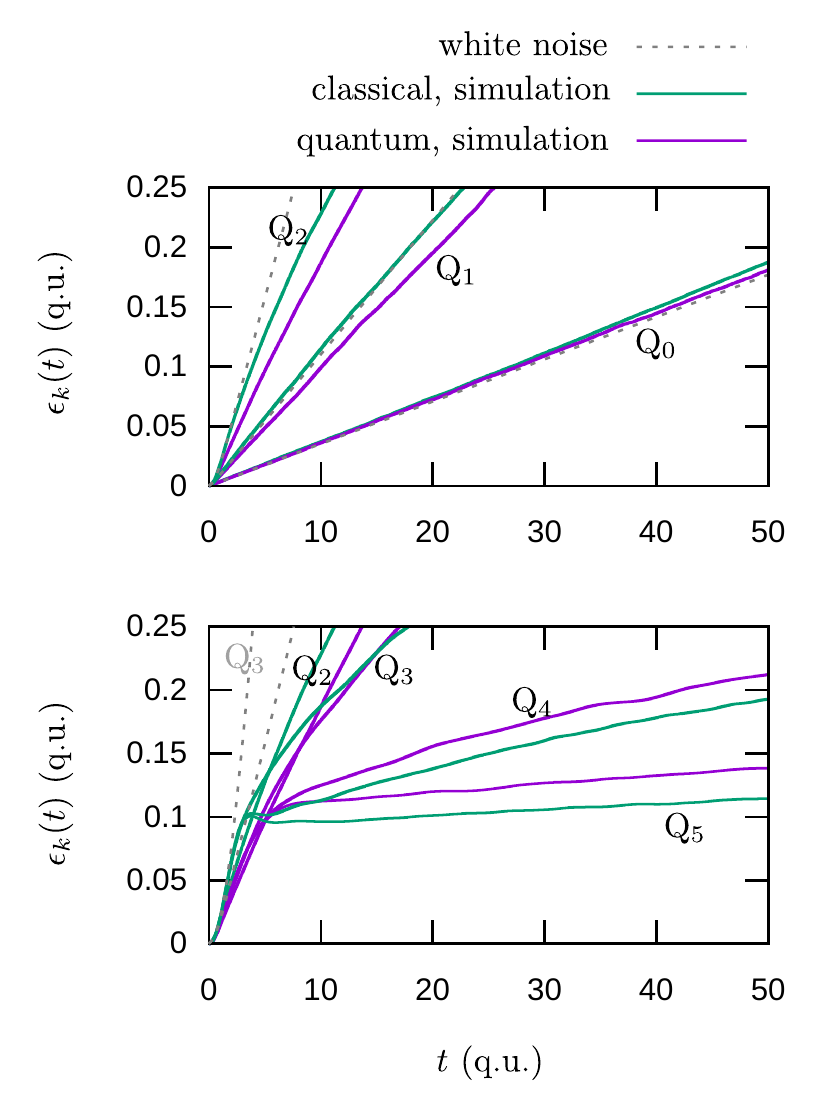}
        \caption{Dependence of average kinetic energy on time for the
        series of ``quantum'' points Q$_n$ in the parameter space.
        Similarly to the B$_n$ points, the behavior changes qualitatively for
        $n=2$, but now the classical and quantum simulations do not agree quantitatively.
	\label{fig-ek-Q} }
\end{figure}

\begin{figure}[t]
        \includegraphics[width=.5\textwidth]{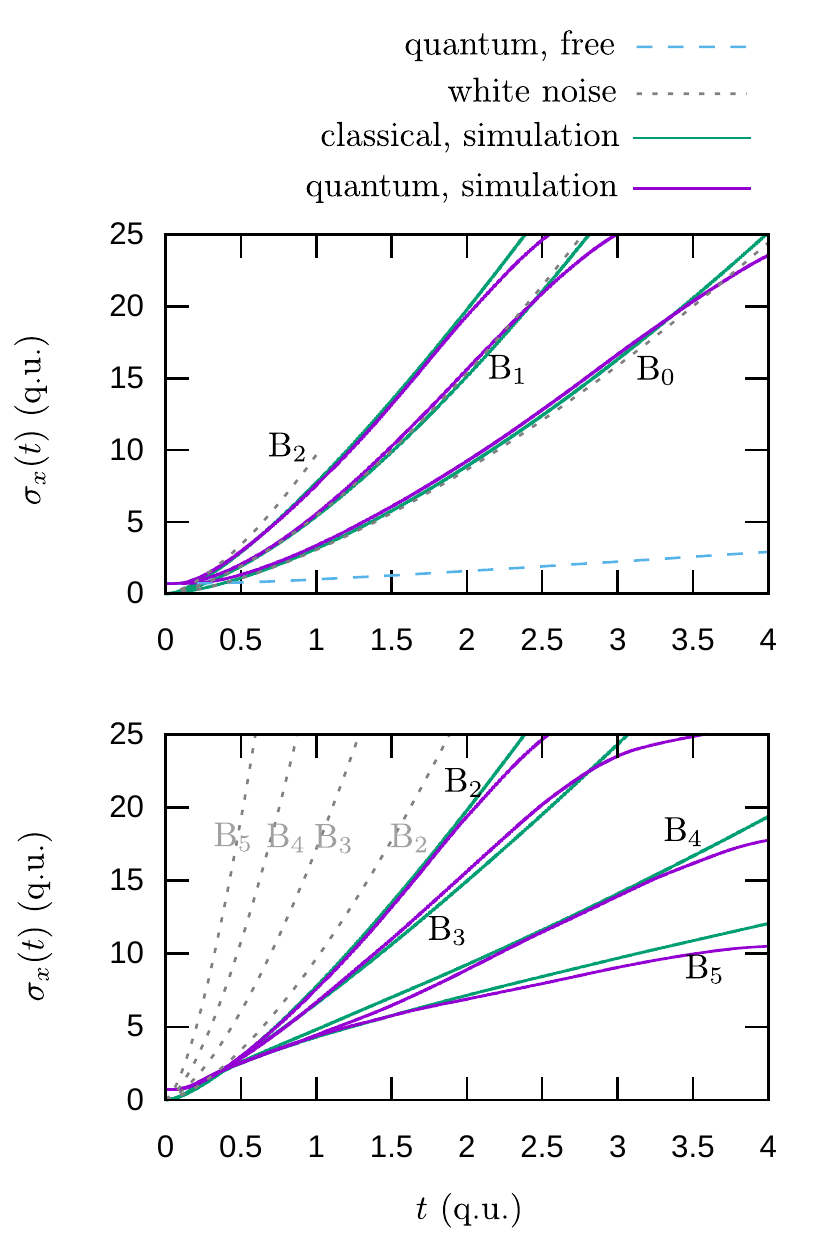}
	\caption{Dependence of the root mean square displacement on time in the classical region of the parameter space. 
	Close to $n=2$ the numerical solutions lead to qualitatively different results from the white noise model.
	\label{fig-sigma-B} }
\end{figure}

\begin{figure}[t]
        \includegraphics[width=.5\textwidth]{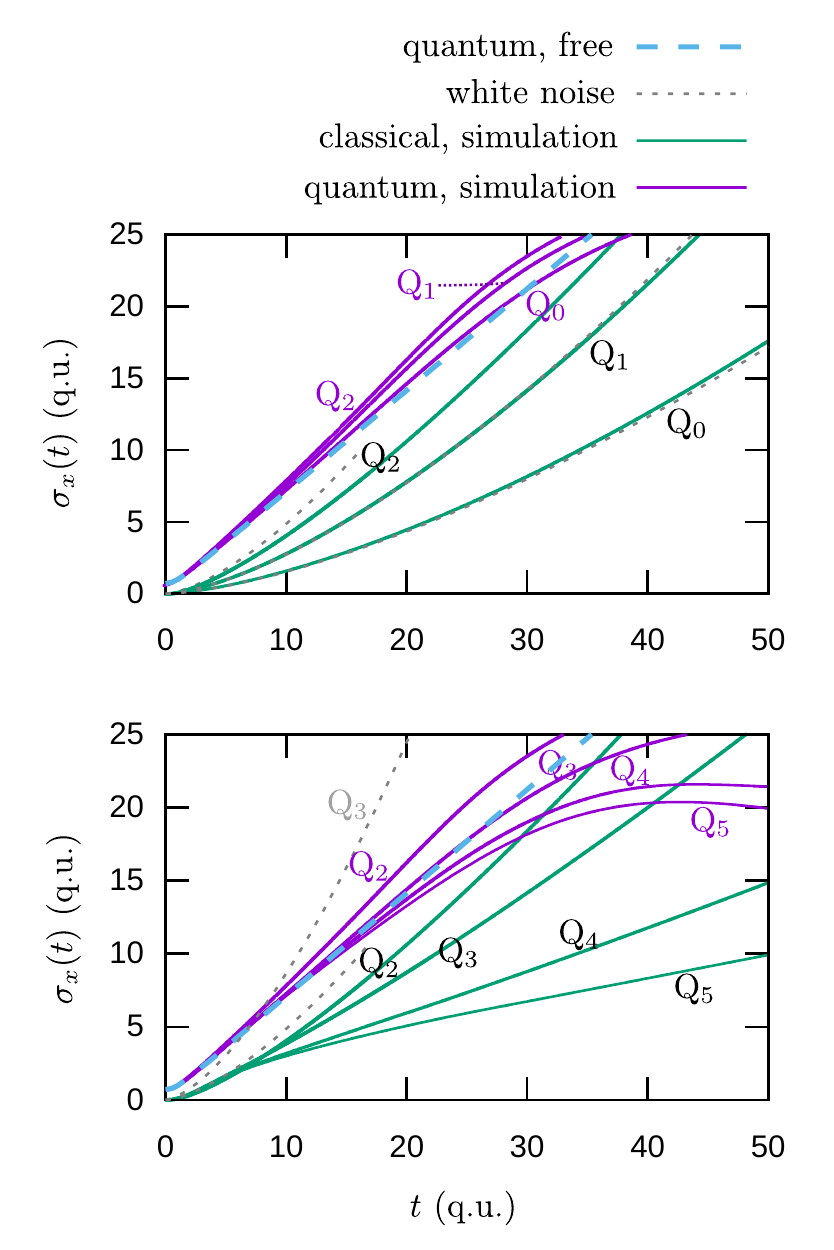}
	\caption{Dependence of the root mean square displacement on time in the quantum region of the parameter space. 
	The behavior of the quantum simulation is dominated by the quantum dispersion of a free particle. \label{fig-sigma-Q} }
\end{figure}

In Figs~\ref{fig-ek-B}, \ref{fig-ek-Q}, \ref{fig-sigma-B} and \ref{fig-sigma-Q} we quantify the dynamics of classical
and quantum simulations for two considered sequences of points B$_n$ ($v_0=50$) and Q$_n$ ($v_0 = 0.2$) from the parameter space.
Both sequences of points correspond to an identical sequence of the parameter
$\tilde v_0 = 10^{n-3}, n = 0,\ldots,5$ and differ only in the importance of
quantum effects: B$_n$ are well in the classical regime where as Q$_n$ increase their quantum
character along the sequence (see also Fig.~\ref{fig-parameter-space}). This choice is motivated by the fact that 
for the two experimental studies \cite{Topinka2001,Patsyk2020} discussed at the end of Sec.~\ref{sec-quantum-model} 
the branching flow pattern
has been observed for $\tilde v_0 \sim 0.05$, i.e. in between the points with $n=1$ and $n=2$.

First we discuss the behavior of the average kinetic energy shown in Figs.~\ref{fig-ek-B} and \ref{fig-ek-Q}.
Whereas for B$_0$ and B$_1$ the average kinetic energy agrees well with the prediction of the white noise model within
the studied time interval, for points B$_n, n \ge 2$ ($\tilde v_0 \ge 0.1$) it attains much lower values,
exhibits concave character and is qualitatively different from the white noise model: by increasing $\tilde v_0$ the
rate of growth of the kinetic energy decreases. Still, for a given $\tilde v_0$ the kinetic energy steadily 
grows with time and at some
point it reaches the characteristic magnitude of potential fluctuation $v_0$ so that the characteristic
time $t_\mathrm{e}$ can be unambiguously determined from the equation $\epsilon_k(t_\mathrm{e}) = v_0$.
For all the B$_n$ points the quantum and the classical simulations give very similar results
since here $v_0 \gg 1$, i.e. one of the two possible conditions discussed in Sec.~\ref{sec-quantum-model} for classical behaviour is fulfilled. 
The above described qualitative change in the behavior of the average kinetic energy
exhibits the model in quantum regime as well, as it is demonstrated in Fig.~\ref{fig-ek-Q}, except that
now the classical and quantum simulations are not in quantitative agreement (see also Fig~\ref{fig-parameter-space}).

Similarly, the root mean square displacements depart from the white-noise model as we move towards
larger $\tilde v_0$ in the parameter space.
In Fig.~\ref{fig-sigma-B}, for B$_n$, $n=0,1,2$ we see a good agreement between the white noise model and both classical 
and quantum simulations, and as of $n=3$ the behavior between the white noise model and the two simulations is qualitatively 
different.
For Q$_n$ points (Fig.~\ref{fig-sigma-Q}) the root mean square displacement obtained from the quantum simulations 
is affected by the uncertainty of the initial state -- all the curves are very close to the free wave packet dispersion
$\sigma^{\mathrm{free}}_{x,\mathrm{quant}}(t)  = \sqrt{ (1 + t^2)/2}$  which is the dominant mechanism for its growth. For larger times
the curves for the quantum simulations bend downwards due to periodic boundary conditions.
Nonetheless it is clear that for this portion of the parameter space
the increase in the dispersion is not due to the fluctuating random potential but
due to the inherent uncertainty of particle velocity set in its initial condition.

\begin{figure}[t]
\includegraphics[width=.5\textwidth]{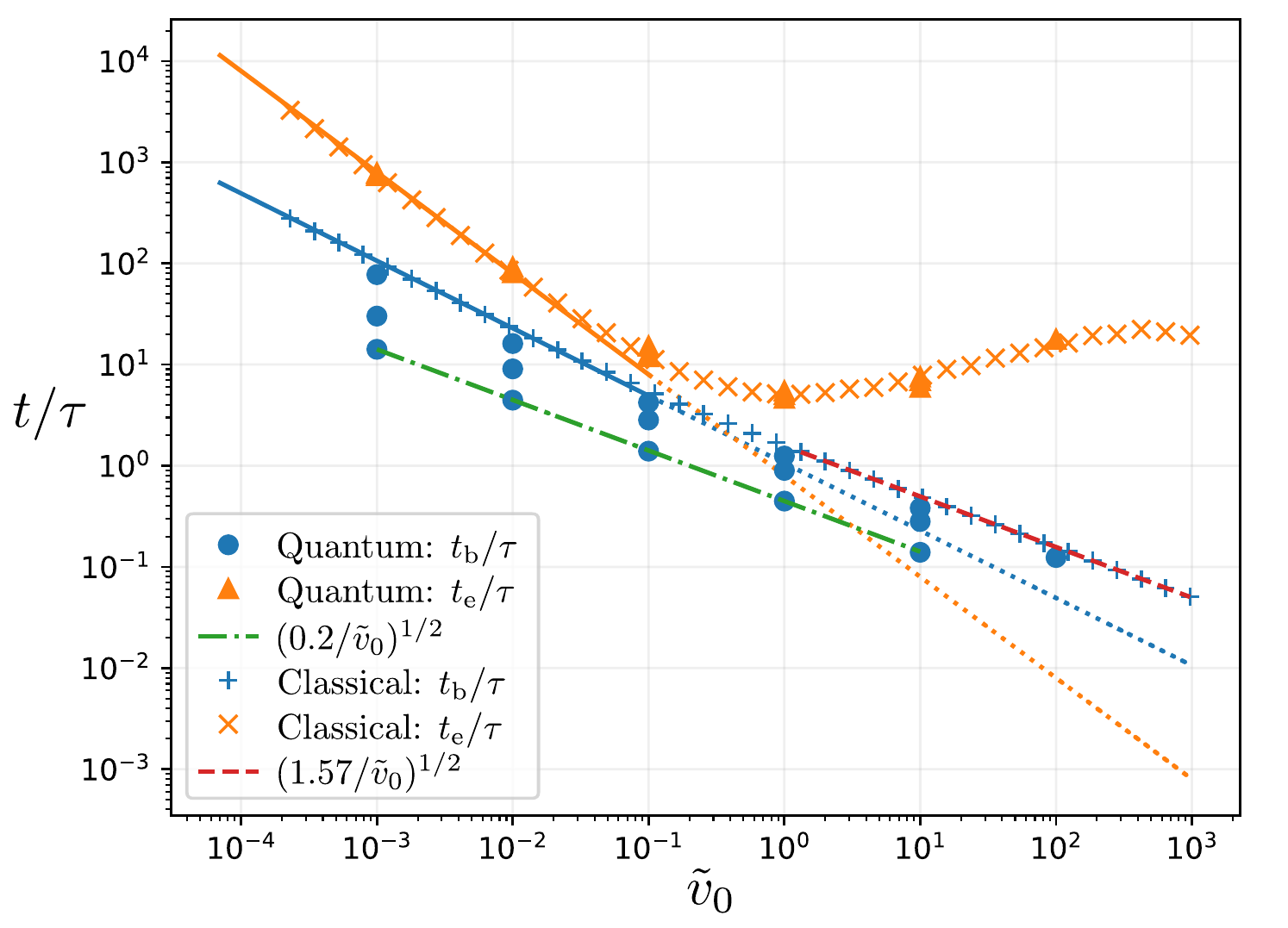}
\caption{
For $\tilde v_0 \ll 1$ the white noise model prediction for both time scales $t_\mathrm{b}
\sim \tilde v_0^{\beta}, \beta = -3/2$ and
$t_\mathrm{e} \sim \tilde v_0^{\epsilon}, \epsilon = -1$ (full lines)
agrees with the classical numerical simulations. At the edge of its validity, $\tilde v_0 \sim 1$
the two time scales approach each other, but eventually separate. The branching time follows a new power
law characterized by exponents $\beta = -1/2$ and the energy time scale attains a well defined minimum.
When quantum effects are important the branching time is not a universal function of
$\tilde v_0$; the vertically stacked circles correspond to identical $\tilde v_0$ and differ in $v_0$. 
In our simulations $t_\mathrm b$ in quantum regime is dominated by wave packet quantum dispersion which is also characterized by 
the exponent $\beta = -1/2$. In contrast, the kinetic energy time scale $t_\mathrm e$, obtained from the plane wave initial condition, 
keeps following the classical behavior.  \label{fig-times} }
\end{figure}

From the above described simulations we obtained the numerically determined times $t_\mathrm{b,class}, t_\mathrm{b,quant}$ and
$t_{e,\mathrm{class}}, t_{e,\mathrm{quant}}$ directly from their definitions $\sigma^2_x(t_\mathrm{b}) = 1$ and
$\epsilon_k(t_\mathrm{e}) = v_0$ respectively. In the region of the phase space where the classical model is valid
we expect that these time scales, when expressed in units of $\tau$, depend solely on a single parameter $\tilde v_0 = v_0 \tau^2$.
The resulting dependences are shown in Fig.~\ref{fig-times}. For $\tilde v_0 < 1$ the white noise model prediction is in agreement with the
numerical simulation with the fluctuating random potential.  The branching time $t_\mathrm b / \tau$ keeps decreasing even for
$\tilde v_0 > 1$ but with a different scaling exponent $t_\mathrm b / \tau \sim \tilde v_0^\beta$, $\beta = -1/2$.
This can be understood as a short transient process in an essentially time-independent random potential ($t \ll \tau$,
i.e. particles' energy is nearly conserved) when the initially homogeneously distributed positions of classical particles with zero kinetic energy
start to move towards the local minima of potential energy. The time to traverse a single correlation length $\lambda = 1$ can be
then estimated from a classical formula obtained from Eqs.~\ref{eq-newton} and the energy conservation,
\beq
        \frac{t_\mathrm b}{\tau} = \int_{x_0}^{x_0+1} \frac{\dd x}{\sqrt{2 \tilde v_0 (\xi(x_0,0) - \xi(x,0))}}
        \sim  \tilde v_0^{-1/2}
\eeq
where $x_0$ is an initial position of a particle (in units of $\lambda$).

The kinetic energy timescale $t_\mathrm{e}$ attains a minimum value for $\tilde v_0 \approx 1.0$ and for larger strengths of the
potential grows again, but the growth is not monotonic. We did not explore its behavior further as our 
interest was only in parameter values few order-of-magnitudes around the experimentally observed systems with branching flow and,
at the same time, the numerical simulations for higher values of $\tilde v_0$ were becoming more demanding.

The time scales need not depend on a single parameter $\tilde v_0 = v_0 \tau^2$ for points in parameter space where 
quantum effect are important. We have seen this behavior already in Fig.~\ref{fig-parameter-space} and 
it can also be manifestly demonstrated in the limit of a free quantum particle $v_0 \rightarrow 0$,
$t_{\mathrm{b,quant}}^{\mathrm{free}}/\tau = 1/\tau = \sqrt{v_0/\tilde v_0}$. The dependence of the branching time for the Q$_n$
points in Fig.~\ref{fig-times} is clearly of this form, with exponent $\beta = -1/2$ and a constant shift in a logarithmic scale that depends on $v_0$.
Obviously, for a different choice of the initial Gaussian wave packet, the dominance 
of the quantum wave packet dispersion appears for different regions of the parameter space.
On the other hand, $t_\mathrm{e, quant}$ stays close to the classical results even for the points in parameter space 
where quantum effect are relevant. In this case we are following the evolution of a plane wave with practically 
zero initial kinetic energy. In the case of initial condition in a form of Gaussian, 
the initial kinetic energy is nonzero ($\epsilon_{k,\mathrm{quant}}(0) = 1/4$ in our simulations).
Specifically for Q$_n$ points this is above $v_0 = 0.2$ so this time scale would be meaningless. 

\section{Conclusions}

In our work we have explored the phenomenon of branching flow for quantum and classical particle moving in a short-ranged 
randomly fluctuating potential in 1D and the validity of the classical white-noise model for its 
description. For this purpose we have introduced a new energy time scale $t_\mathrm{e}$ that complements the branching time 
$t_\mathrm{b}$. In the classical regime the behavior of these two 
time scales can be well described within a white noise model for the random force as long as its strength parameter $\tilde v_0 \leq 0.1$. 
On the other hand, for $\tilde v_0 > 1.0$ the branching time is described by a different scaling exponent for which we have 
provided a simple physical explanation in terms of a short time dynamics in a vicinity of a local minimum in the random potential. 
The energy time scale $t_\mathrm{e}$ attain a minimum for $\tilde v_0 \approx 1$ and grows 
nonmonotically beyond this value for $\tilde v_0 > 1.0$. 
Interestingly, for the values of $\tilde v_0 \sim 0.01$ -- $0.1$, where branching flow has been experimentally 
observed in the past, these two time scales are closest to each other. 
This can be understood by ruling out
the other possibilities as follows: for $\tilde v_0 > 1.0$ the branching pattern disappears due to stronger localisation of the particles/waves
and for $\tilde v_0 \ll 1.0$ the $t_\mathrm e$ is too large so that it takes too much time to develop a widely branched pattern.
Hence, for experimental observation it is preferable to have as small $t_\mathrm e$ as possible, as long as the localisation does not take over. 

We have discussed that the importance of quantum effects is accompanied by a loss of simple dependence of system's behavior 
on a single parameter $\tilde v_0 = v_0 \tau^2$; instead both the random potential root mean square amplitude $v_0$ and 
its correlation time $\tau$ have to be considered as independent parameters. The quantum effects are important if two conditions 
$v_0 < 1$ and $\tau > 1$ are fulfilled at once, which we have expected based on dimensional arguments and confirmed by extensive 
numerical simulations.

\end{document}